\documentstyle[12pt]{article}

\catcode`@=11 \@addtoreset{equation}{section} \catcode`@=12

\begin{document}
\begin{titlepage}
\begin{flushright}\vbox{\begin{tabular}{c}
           TIFR/TH/97-32\\
           June, 1997\\
           hep-ph/9706541\\
\end{tabular}}\end{flushright}
\begin{center}
   {\large \bf
      $\chi_1$ and Polarisation Asymmetries for Quarkonia\\
      at High Orders in Non-relativistic QCD}
\end{center}
\bigskip
\begin{center}
   {Sourendu Gupta\footnote{E-mail: sgupta@theory.tifr.res.in}
    and Prakash Mathews\footnote{E-mail: prakash@theory.tifr.res.in}.\\
    Theory Group, Tata Institute of Fundamental Research,\\
    Homi Bhabha Road, Bombay 400005, India.}
\end{center}
\bigskip
\begin{abstract}
We study doubly polarised asymmetries of $\bar cc$ and $\bar bb$ mesons 
in hadro- and photo-production at low transverse momentum in 
non-relativistic QCD to high orders in the relative velocity of the pair,
$v$. We give the complete set of expressions required for the asymmetries
up to order $v^9$. The asymmetries in the production of $\eta_{c,b}$ states
are a stable measure of the polarised gluon densities. The asymmetries for
$\chi_{c,b}$, $J/\psi$, $\psi'$, and the various $\Upsilon$ states are
stringent tests of the NRQCD scaling relations.
\end{abstract}
\end{titlepage}

\section{\label{intro}Introduction}

Recent progress in the understanding of cross sections for production of
heavy quarkonium resonances has come through the non-relativistic QCD (NRQCD)
reformulation of this problem \cite{caswell}. Factorisation of the physics
at a short distance scale of order $1/m$ (where $m$ is the mass of heavy quark)
from long distance scales of order $1/mv$ and $1/mv^2$ (where $v$ is the
velocity of either of the heavy quark in the rest frame of the pair) has been
proven in the NRQCD formalism for processes dominated by a large transverse
momentum \cite{bbl}. The resulting cross sections are a double power series
in $v^2$ and the strong coupling 
$\alpha_{\scriptscriptstyle S}(m^2)$.
For charmonium states, a numerical coincidence,
$v^2\sim\alpha_{\scriptscriptstyle S}(m^2)$, complicates the double expansion.

The formalism has been successfully applied to large transverse momentum
processes \cite{jpsi}. Interestingly, inclusive production cross sections
for charmonium at low energies, dominated by low transverse momenta, also
seem to have a good phenomenological description in terms of this approach
\cite{ours,br,our2}. The remaining phenomenological problem is to explain
the $\chi_1/\chi_2$ ratio observed in hadro-production. $\chi_{0,2}$ are first
produced at order $\alpha^2_{\scriptscriptstyle S}v^5$, whereas $\chi_1$
begins only at order $\alpha^2_{\scriptscriptstyle S}v^9$. It was argued
in \cite{br,our2} that a better understanding of this ratio may require
higher order terms in $\alpha_{\scriptscriptstyle S}$ and $v^2$. In
\cite{upol} the NRQCD series was evaluated to order
$\alpha^2_{\scriptscriptstyle S}v^9$. In the dominant 
gluon fusion channel
a previously missed term at order $\alpha^2_{\scriptscriptstyle S}v^7$
was found for $\chi_{0,2}$ and up to eleven terms at order
$\alpha^2_{\scriptscriptstyle S}v^9$ were obtained for all $\chi$ states.

In this paper we extend the computations of \cite{upol} to polarisation
asymmetries,
\begin{equation}
  A \;=\; {\Delta\sigma\over\sigma} \;=\;
    {\sum_{hh'} hh'\sigma(h,h')\over\sum_{hh'} \sigma(h,h')},
\label{intro.asym}\end{equation}
where $h$ and $h'$ are the helicities of the beam and target respectively,
and $\sigma(h,h')$ is the cross section for fixed initial helicities. This
allows us to formulate several tests of NRQCD and the velocity scaling
rules that it yields. In addition, we
propose measurements of polarised gluon densities, $\Delta g$. The spin
asymmetries have earlier been computed at lower orders for low transverse
momentum processes \cite{pol} and at high transverse momenta \cite{polo}.

The NRQCD factorisation formula for the inclusive production of heavy
quarkonium resonances $H$ with 4-momentum $P$ can be written as
\begin{equation}\begin{array}{rl}
   d\sigma\;=\;&
      {\displaystyle{1\over\Phi}{d^3P\over(2\pi)^3 2E_{\scriptscriptstyle P}}}
     \sum_{ij} C_{ij}
          \left\langle{\cal K}_i\Pi(H){\cal K}^\dagger_j\right\rangle,\\
   d\Delta\sigma\;=\;&
      {\displaystyle{1\over\Phi}{d^3P\over(2\pi)^3 2E_{\scriptscriptstyle P}}}
       \sum_{ij} {\widetilde C}_{ij}
          \left\langle{\cal K}_i\Pi(H){\cal K}^\dagger_j\right\rangle.
\end{array}\label{intro.nrqcd}\end{equation}
where $\Phi$ is a flux factor. The coefficient functions $C_{ij}$ and
${\widetilde C}_{ij}$ are computable in perturbative QCD and hence have an
expansion in the strong coupling $\alpha_{\scriptscriptstyle S}$ (evaluated
at the NRQCD cutoff). Although the matrix element is non-perturbative, it has
a fixed scaling dimension in the quark velocity $v$.

The fermion bilinear operators ${\cal K}_i$ are built out of heavy quark
fields sandwiching colour and spin matrices and the covariant derivative
${\bf D}$. The composite labels $i$ and $j$ include the colour index $\alpha$,
the spin quantum number $S$, the number of derivative operators $N$, the
orbital angular momentum $L$ (found by coupling the derivatives), the total
angular momentum $J$ and the helicity $J_z$. When the final state spin is
unobserved, the hadron projection operator 
\begin{equation}
   \Pi(H)\;=\;{\sum_s} \left|H,s\rangle\langle H,s\right|,
\label{intro.hproj}\end{equation}
(where $s$ denotes hadron states with energy less than the NRQCD cutoff),
is diagonal in these quantum numbers. Then it is clear that the operators
${\cal K}_i$ and ${\cal K}_j$ in eq.\ (\ref{intro.nrqcd}) are restricted to
have equal $L$, $S$, $J$ and $J_z$, for both $d\sigma$ and $d\Delta\sigma$.

The $J_z$-dependence of these matrix elements can be factored out using the
Wigner-Eckart theorem---
\begin{equation}\begin{array}{rl}
   \langle{\cal K}_i\Pi(H){\cal K}^\dagger_i\rangle\;=\;&
       {\displaystyle{1\over2J+1}} {\cal O}^H_\alpha({}^{2S+1}L_J^N),\\
   \langle{\cal K}_i\Pi(H){\cal K}^\dagger_j\rangle\;=\;&
       {\displaystyle{1\over2J+1}}
                 {\cal P}^H_\alpha({}^{2S+1}L_J^N,{}^{2S+1}L_J^{N'}),
\end{array}\label{intro.ored}\end{equation}
where the factors of $1/(2J+1)$ come from a
Clebsch-Gordan coefficient. This is conventionally included in the coefficient
function. We have used the notation of \cite{bbl} for both diagonal and
off-diagonal operators. The NRQCD power counting rule for these matrix elements
is---
\begin{equation}
   d\;=\;3+N+N'+2(E_d+2M_d),
\label{intro.rule}\end{equation}
where $E_d$ and $M_d$ are the number of colour electric and magnetic
transitions required to connect the hadronic state to the state
${\cal K}_i|0\rangle$.

In section \ref{te} of this paper we present the main results for
$\Delta\sigma$ after a telegraphic review of the threshold expansion technique
\cite{bchen}, the kinematics and the appropriate Taylor series expansion of
the perturbative matrix element. This material is given in greater detail in
our earlier paper \cite{upol}. The phenomenology is discussed in the final
section \ref{disc}.

\section{\label{te}The Threshold Expansion}

We choose to construct the coefficient functions using the ``threshold
expansion'' technique of \cite{bchen}. This consists of calculating, in
perturbative QCD, the matrix element ${\cal M}$ connecting the initial
states to final states with a heavy quark-antiquark pair ($\bar QQ$),
and Taylor expanding the result in the relative momentum of the pair,
$q$, after performing a non-relativistic reduction of the Dirac spinors.
The resulting expression is squared and matched to the NRQCD formula of
eq.\ (\ref{intro.nrqcd}) by inserting a perturbative projector onto a
non-relativistic $\bar QQ$ state between the two spinor bilinears. The
coefficient of this matrix element is the required coefficient function.

Symbolically---
\begin{equation}\begin{array}{rl}
   {1\over4}\sum_{hh'} |{\cal M}(h,h')|^2 \;=\;& \sum_{ij} C_{ij}
      \left\langle{\cal K}_i\Pi(\bar QQ){\cal K}^\dagger_j\right\rangle,\\
   {1\over4}\sum_{hh'} hh'|{\cal M}(h,h')|^2 \;=\;& \sum_{ij} \widetilde C_{ij}
      \left\langle{\cal K}_i\Pi(\bar QQ){\cal K}^\dagger_j\right\rangle,
\end{array}\label{te.matching}\end{equation}
where the left hand sides are Taylor expanded in $q$. Each factor of $q$
Fourier transforms into a factor of the covariant derivative ${\bf D}$ on
the right hand side. Since each matrix element on the right of
eq.\ (\ref{te.matching}) corresponds to an unique matrix element in
eq.\ (\ref{intro.nrqcd}), the order up to which the Taylor expansion is to be
performed is determined by the scaling of the non-perturbative matrix
elements with $v$. Since we require a classification of operators by the
angular momentum, it turns out to be very convenient to use the spherical
tensor methods detailed in \cite{upol}.

In this paper we evaluate the polarised cross sections to order
$\alpha^2_{\scriptscriptstyle S}v^9$. The Taylor expansion order, $N+N'\le6$
is obtained by setting $d=9$ and $E_d=M_d=0$ in eq.\ (\ref{intro.rule}).
Furthermore, since we examine the leading term in perturbation theory, the
perturbative projector has only one term---
\begin{equation}
   \Pi(\bar QQ)\;=\;|\bar QQ\rangle\langle\bar QQ|.
\label{te.qqbproj}\end{equation}
In agreement with \cite{upol,bchen} we use the relativistic normalisation
of states
\begin{equation}
   \langle Q(p,\xi)\bar Q(q,\eta) | Q(p',\xi')\bar Q(q',\eta')\rangle
     \;=\;4 E_p E_q (2\pi)^6\delta^3(p-p')\delta^3(q-q'),
\label{te.norm}\end{equation}
with the spinor normalisations $\xi^\dagger\xi=\eta^\dagger\eta=1$. Expanding
$E_p=E_q=\sqrt{m^2+q^2}$ in $q^2$ allows us to write the spinor bilinears in
terms of transition operators built out of the heavy quark field. The
kinematics is very simple to leading order in $\alpha_{\scriptscriptstyle S}$.
The momenta of the initial particles are $p_1$ and $p_2$. We take $p_1$ to lie
in the positive $z$-direction and $p_2$ to be oppositely directed. The net
momentum $P=p_1+p_2$.

\begin{figure}
\vskip7truecm
\includegraphics{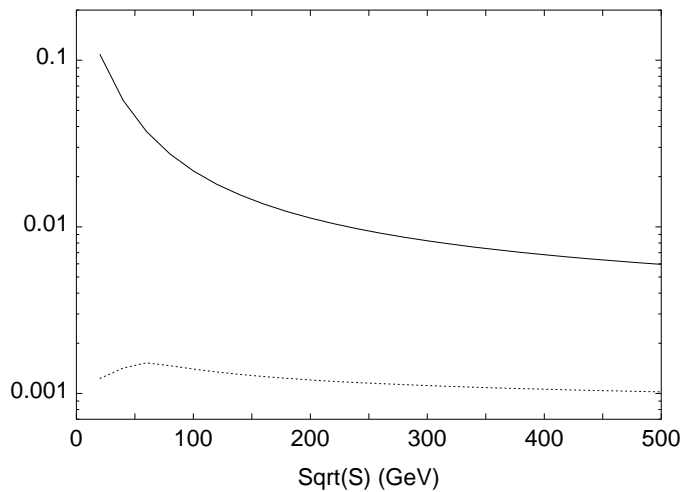}
\caption[dummy]{The ratios ${\cal L}_{\bar qq}/{\cal L}_{gg}$ (full line)
  and $\Delta{\cal L}_{\bar qq}/{\cal L}_{gg}$ (broken line) computed
  using the GRSV LO parton density set \cite{grv} as a function of the
  CM energy $\sqrt S$ at zero rapidity.}
\label{fg.lumin}\end{figure}

The 4-momenta of $Q$ and $\bar Q$ ($p$ and $\bar p$ respectively) are
written as
\begin{equation}
   p\;=\;{1\over2}P+L_j q^j\qquad{\rm and}\qquad
   \bar p\;=\;{1\over2}P-L_j q^j.
\label{me.momdef}\end{equation}
Note that $p^2=\bar p^2=m^2$, where $m$ is the mass of the heavy quark.
The space-like vector $q$ is always defined in the rest frame of the pair,
and $L^\mu_j$ boosts it to any frame. We shall use Greek indices for Lorentz
tensors and Latin indices for Euclidean 3-tensors.

In the next two sections we will write down the parton level cross sections,
$\Delta\hat\sigma$ for $\bar qq\to H$ and $gg\to H$ (where $H$ is a
quarkonium state). The hadronic cross sections, $\Delta\sigma$, are obtained
by multiplying these by appropriate parton luminosities---
\begin{equation}
   \Delta{\cal L}_{\bar qq}\;=\;\sum_f\Delta q_f(x_1)\Delta\bar q_f(x_2),\quad
   \Delta{\cal L}_{gg}\;=\;\Delta g(x_1)\Delta g(x_2),
\label{me.lumin}\end{equation}
where $x_1=(2m/\sqrt S)\exp(y)$ and $x_2=(2m/\sqrt S)\exp(-y)$, $\sqrt S$
is the centre of mass (CM) energy of the colliding protons, $y$ is the rapidity
at which the quarkonium is observed, and $\Delta q_f(x)$, $\Delta \bar q_f(x)$
and $\Delta g(x)$ are the quark, antiquark (of flavour $f$) and gluon polarised
densities. For unpolarised cross sections, $\sigma$, we must similarly define
unpolarised luminosities ${\cal L}_{\bar qq}$ and ${\cal L}_{gg}$ using the
corresponding unpolarised parton densities. As shown in Figure \ref{fg.lumin},
for $\sqrt S\ge20$ GeV, ${\cal L}_{\bar qq}\ll{\cal L}_{gg}$. Also, present
day data indicates that $|\Delta{\cal L}_{\bar qq}|\ll{\cal L}_{gg}$.
Consequently, the $\bar qq$ channel may be neglected for double polarised
asymmetries to good precision.

\subsection{$\bar qq\to\bar QQ$}

We write down the polarised matrix element for the subprocess $\bar qq\to\bar
QQ$ for completeness, and because it provides a simple illustration of the
techniques used. It is given exactly by the expression
\begin{equation}\begin{array}{rl}
   {\cal M}\;&=\;-{\displaystyle{ig^2\over M^2}}
      \left[\bar v(p_2,h^\prime)\gamma_\mu T^a u(p_1,h)\right] L^\mu_j
     \\ & \qquad\quad \times
    \left[M\xi^\dagger\sigma^jT^a\eta - 
        {\displaystyle{4\over M+2m}}
      \,q^j\xi^\dagger(q\cdot\sigma)T^a\eta\right],
\end{array}\label{me.qqmat}\end{equation}
where $T^a$ is a colour generator, $u$ and $v$ are the light quark spinors 
with helicities $h$ and $h^\prime$ respectively, and $\xi$ and $\eta$ are
the heavy quark Pauli spinors. 
The equations of motion for the initial state quarks have been used to obtain 
the explicitly gauge invariant matrix element above. The 
desired Taylor series expansion is obtained by using the relation $M^2=4
(m^2+q^2)$ to expand all factors with $M$. 

The squared matrix element for this process is easily written down.
The difference between the polarised and unpolarised cross sections follows
when one notices that the light quark helicities give an overall factor of
$1-hh'$ in the squared matrix element. As a result, $\widetilde C_{ij}=-C_{ij}$
and the parton level spin asymmetries are all $-1$. We list the parton level
polarised cross sections---
\begin{equation}\begin{array}{rl}
   \Delta \hat\sigma^{\eta_c}_{\bar qq} \;&=\;
       - \displaystyle{\pi^3\alpha_s^2\over54m^4}
          \delta(\hat s-4m^2){\cal O}^{\eta_c}_8({}^3 S^0_1),\\

   \Delta \hat\sigma^{h_c}_{\bar qq} \;&=\;
       - \displaystyle{\pi^3\alpha_s^2\over54m^4}
          \delta(\hat s-4m^2){\cal O}^{h_c}_8({}^3 S^0_1),\\

   \Delta \hat\sigma^{J/\psi}_{\bar qq} \;&=\;
       - \displaystyle{\pi^3\alpha_s^2\over54 m^4}
          \delta(\hat s-4m^2)\biggl[{\cal O}^{J/\psi}_8({}^3 S^0_1)
       \\&\qquad\qquad
                +\displaystyle{1\over m^2}\biggl\{
                   \displaystyle{2\over\sqrt3}
                      {\cal P}^{J/\psi}_8({}^3 S^0_1,{}^3 S^2_1)
                  +\displaystyle{1\over4}
                      {\cal O}^{J/\psi}_8({}^3 P^2_1)\biggr\}\biggr],\\

   \Delta \hat\sigma^{\chi_J}_{\bar qq} \;&=\;
       - \displaystyle{\pi^3\alpha_s^2\over54m^4}
          \delta(\hat s-4m^2)\biggl[{\cal O}^{\chi_J}_8({}^3 S^0_1)
                +\displaystyle{2\over\sqrt3m^2}
                     {\cal P}^{\chi_J}_8({}^3 S^0_1,{}^3 S^2_1)
       \\&\qquad
            +\displaystyle{1\over m^4}\left\{
                 \displaystyle{4\over3}{\cal O}^{\chi_J}_8({}^3 S^2_1)
                +\displaystyle{5\over12}{\cal O}^{\chi_J}_8({}^3 D^2_1)
                +\displaystyle{7\sqrt5\over12}
                     {\cal P}^{\chi_J}_8 ({}^3 S^0_1,{}^3 S^4_1)\right\}
          \biggr],
\end{array}\label{me.xsecq}\end{equation}
where $\hat s$ is the parton CM energy.

\subsection{$gg\to\bar QQ$}

The squared matrix element for the $gg$ process is technically a little
more complicated. We work in a class of ghost-free gauges called planar
gauges \cite{ddt}. The density matrix for an initial state gluon of
momentum $p$ and helicity $h$ in these gauges is given by
\begin{equation}
\epsilon_\mu(p,h)\epsilon_\nu^*(p,h)\;=\; {1\over2}
     \left[ -g_{\mu\nu} + {1\over p\cdot V} (p_\mu V_\nu+p_\nu V_\mu)
     + {i h\over p\cdot V} \epsilon_{\mu \nu \rho \sigma} p^\rho V^\sigma 
     \right]. 
\label{me.gauge}\end{equation}
The vector $V$ defines the gauge choice. We write $V=c_1 p_1+c_2 p_2$, with 
$c_1/c_2\sim{\cal O}(1)$. We verify that all our results are gauge invariant
by the explicit check that they do not depend on the arbitrary coefficients 
$c_1$ and $c_2$.

The sum of the matrix elements arising from the three Feynman diagrams
($s$-channel gluon exchange, ${\cal M}_s$, and $t$- and $u$-channel
quark exchanges, ${\cal M}_t$ and ${\cal M}_u$) can be decomposed into
three colour amplitudes---
\begin{equation}
   {\cal M}\;=\;{1\over6}g^2\delta_{ab}S
               +{1\over2}g^2d_{abc}D^c
               +{i\over2}g^2f_{abc}F^c.
\label{me.ampl}\end{equation}
The colour amplitudes $S$ and $D$ involve only ${\cal M}_t+{\cal M}_u$,
whereas $F$ involves ${\cal M}_s$ as well as ${\cal M}_t-{\cal M}_u$. 

In order to write down our results, we find it convenient to introduce the
notation
\begin{equation}
  {\cal A}\;=\;{1\over M^2}\varepsilon_{\lambda\sigma\mu\nu}
         p_1^\lambda p_2^\sigma\epsilon_1^\mu\epsilon_2^\nu
     \qquad{\rm and}\qquad
  {\cal S}_{ij}\;=\;A_i\hat z_j+A_j\hat z_i-B_{ij}
               +\epsilon_1\cdot\epsilon_2\hat z_i\hat z_j,
\label{me.not4}\end{equation}
where
\begin{equation}\begin{array}{rl}
   A_i\;&=\;{1\over M}\left(\epsilon_1\cdot L_i \epsilon_2\cdot p_1
       -\epsilon_2\cdot L_i \epsilon_1\cdot p_2\right),\\
   B_{ij}\;&=\;\epsilon_1\cdot L_i \epsilon_2\cdot L_j
       +\epsilon_2\cdot L_i \epsilon_1\cdot L_j.
\end{array}\label{me.not2}\end{equation}
Here $\epsilon_i$ is the polarisation vector for the initial gluon of momentum
$p_i$.
In order to identify all terms to order $v^9$ we need the
colour amplitude $S$ to order $q^5$---
\begin{equation}\begin{array}{rl}
   S\;&=\;-\left({\displaystyle8im\over\displaystyle M}\right){\cal A}
                   \,(\xi^\dagger\eta)
       + {\displaystyle 4\over\displaystyle M} {\cal S}_{jm}
                   \,(q^m\xi^\dagger\sigma^j\eta)
       - \left({\displaystyle32im\over\displaystyle M^3}\right){\cal A}
                   \hat z_m\hat z_n\,(q^mq^n\xi^\dagger\eta)
     \\ & \qquad\quad
       + {\displaystyle16\over\displaystyle M^3}
           \left[{\cal S}_{jm}\hat z_n\hat z_p
                     -{\displaystyle M\over\displaystyle M+2m}\delta_{jm}
                {\cal S}_{np}\right]\,(q^mq^nq^p\xi^\dagger\sigma^j\eta)
     \\ & \qquad\qquad
       - \left({\displaystyle128im\over\displaystyle M^5}\right){\cal A}
                   \hat z_m\hat z_n\hat z_p\hat z_r
             \,(q^mq^nq^pq^r\xi^\dagger\eta)
     \\ & \quad\quad
       + {\displaystyle64\over\displaystyle M^5}
           \left[{\cal S}_{jm}\hat z_n\hat z_p
                     -{\displaystyle M\over\displaystyle M+2m}\delta_{jm}
                {\cal S}_{np}\right]\hat z_r\hat z_s
                     \,(q^mq^nq^pq^rq^s\xi^\dagger\sigma^j\eta),
\end{array}\label{me.ampls}\end{equation}
where $\hat z$ is the unit vector in the $z$-direction. The amplitude $D$
differs only through having colour octet matrix elements in place of the
colour singlet ones shown above. For the colour amplitude $F$ we need the
expansion
\begin{equation}
   F^c\;=\;-\left({\displaystyle16im\over\displaystyle M^2}\right){\cal A}
                   \hat z_m\,(q^m\xi^\dagger T^c\eta)
       + {\displaystyle8\over\displaystyle M^2}{\cal S}_{jm}\hat z_n
                   \,(q^mq^n\xi^\dagger\sigma^jT^c\eta).
\label{me.amplf}\end{equation}
In all three colour amplitudes, the terms in ${\cal A}$ are spin singlet and
those in ${\cal S}$ are spin triplet.

The density matrix in eq.\ (\ref{me.gauge}) yields
\begin{equation}\begin{array}{rl}
   \Delta{\cal S}\cdot{\cal S}^*\,\equiv\,&
     {1\over4} \sum_{h h'} h h' S_{jm}S^*_{j'm'}\,=\,
    {3\over2}[\sigma q]^0_0[\sigma^\dagger q^\dagger]^0_0
     - \sum_{\lambda=\pm2}[\sigma q]^2_\lambda
                        [\sigma^\dagger q^\dagger]^2_{-\lambda},\\
   {\cal S}\cdot{\cal S}^*\,\equiv\,&
     {1\over4} \sum_{h h'} S_{jm}S^*_{j'm'}\,=\,
    {3\over2}[\sigma q]^0_0[\sigma^\dagger q^\dagger]^0_0
     + \sum_{\lambda=\pm2}[\sigma q]^2_\lambda
                        [\sigma^\dagger q^\dagger]^2_{-\lambda}.
\end{array}\label{me.ident}\end{equation}
We also find, ${\cal A}\cdot{\cal A}^*=\Delta{\cal A}\cdot{\cal A}^*=1/8$ and
${\cal A}\cdot{\cal S}^*=\Delta{\cal A}\cdot{\cal S}^*=0$.
Consequently, the difference between the unpolarised
\cite{upol} and polarised cases lies solely in the flipped sign of
the $J=2$ part in eq.~(\ref{me.ident}). Furthermore, even and odd terms
in the three colour amplitudes of eq.\ (\ref{me.ampl}) do not interfere
with each other. 

The direct ${}^3S_1$ polarised subprocess cross section is
\begin{equation}
   \Delta \hat\sigma^{J/\psi}_{gg}(\hat s) \;=\;
          \varphi \left[{5\over48}\widetilde\Theta^{J/\psi}_D(7)
                  +\left\{{5\over48}\widetilde\Theta^{J/\psi}_D(9)
                        +{3\over16}\widetilde\Theta^{J/\psi}_F(9)\right\}
		\right],
\label{me.jpsi}\end{equation}
where
\begin{equation}
\varphi = {\pi^3\alpha_s^2\over4m^2}\delta(\hat s-4 m^2),
\label{me.varphi}\end{equation}
and $\widetilde\Theta^{J/\psi}_a(d)$ denotes combinations of non-perturbative
matrix elements from the colour amplitude $a$ ($=S$, $D$ or $F$) at order 
$v^d$ for the polarised subprocess cross section. These are given by
\begin{equation}\begin{array}{rl}
   \widetilde\Theta^{J/\psi}_D(7)\;=\;&
        {\displaystyle{1\over2m^2}}{\cal O}^{J/\psi}_8({}^1S_0^0)
       +{\displaystyle{1\over2m^4}}\left[
               3{\cal O}^{J/\psi}_8({}^3P_0^1)
              -{\displaystyle{4\over5}}{\cal O}^{J/\psi}_8({}^3P_2^1)
                                   \right], \\

   \widetilde\Theta^{J/\psi}_D(9)\;=\;&
        {\displaystyle{1\over\sqrt3m^4}}{\cal P}^{J/\psi}_8({}^1S_0^0,{}^1S_0^2)
       +{\displaystyle{1\over\sqrt{15}m^6}}\biggl[
           {\displaystyle{35\over4}}
                    {\cal P}^{J/\psi}_8({}^3P_0^1,{}^3P_0^3)
      \\&\qquad
          -2 {\cal P}^{J/\psi}_8({}^3P_2^1,{}^3P_2^3)
                                   \biggr], \\

   \widetilde\Theta^{J/\psi}_F(9)\;=\;& \displaystyle{1\over2 m^6} 
               \left[{1\over3} {\cal O}^{J/\psi}_8 ({}^3P^2_1)+
             {2\over5} {\cal O}^{J/\psi}_8 ({}^3P^2_2) \right].\\
\end{array}\label{me.jpsime}\end{equation}
As a consequence of eq.~(\ref{me.ident}) the coefficient of $J=2$ matrix 
elements changes sign between the polarised and unpolarised cases.  
The cross sections for $\psi'$, $\Upsilon$ and all other ${}^3S_1$ states
need only the replacement of the appropriate matrix elements in eq.\ 
(\ref{me.jpsime}).

The ${}^3P_0$ polarised subprocess cross section is
\begin{equation}\begin{array}{rl}
   \Delta \hat\sigma^{\chi_0}_{gg}(\hat s) \;=&\;
	\varphi \biggl[\displaystyle{1\over18}\widetilde\Theta^{\chi_0}_S(5)
             +\displaystyle{1\over18}\widetilde\Theta^{\chi_0}_S(7)
          \\&\qquad\qquad\qquad
             +\left\{\displaystyle{1\over18}\widetilde\Theta^{\chi_0}_S(9) 
             +\displaystyle{5\over48}\widetilde\Theta^{\chi_0}_D(9)
             +\displaystyle{3\over16}\widetilde\Theta^{\chi_0}_F(9)\right\}
             \biggr].
\end{array}\label{me.chi0}\end{equation}
where combinations of non-perturbative matrix elements appearing in 
eq. (\ref{me.chi0}) are
\begin{equation}\begin{array}{rl}
   \widetilde\Theta^{\chi_0}_S(5)\;=\;&
        {\displaystyle{3\over2m^4}}{\cal O}^{\chi_0}_1({}^3P_0^1),\\

   \widetilde\Theta^{\chi_0}_S(7)\;=\;&
        {\displaystyle{7\sqrt5\over4\sqrt3m^6}}
                {\cal P}^{\chi_0}_1({}^3P_0^1,{}^3P_0^3),\\

   \widetilde\Theta^{\chi_0}_S(9)\;=\;&
       {\displaystyle{1\over8m^8}}\biggl[
           {\displaystyle{245\over9}}{\cal O}^{\chi_0}_1({}^3P_0^3)
          +{\displaystyle{149\sqrt7\over10\sqrt3}}
               {\cal P}^{\chi_0}_1({}^3P_0^1,{}^3P_0^5)
                                  \biggr]
       -{\displaystyle{2\over5m^4}}{\cal O}^{\chi_0}_1({}^3P_2^1),\\

   \widetilde\Theta^{\chi_0}_D(9)\;=\;&
        {\displaystyle{1\over2m^2}}{\cal O}^{\chi_0}_8({}^1S_0^0)
       +{\displaystyle{1\over2m^4}}\left[
               3{\cal O}^{\chi_0}_8({}^3P_0^1)
              -{\displaystyle{4\over5}}{\cal O}^{\chi_0}_8({}^3P_2^1)
                                   \right], \\

   \widetilde\Theta^{\chi_0}_F(9)\;=\;&
           \displaystyle{1\over 6 m^4}{\cal O}^{\chi_0}_8 ({}^1P^1_1)
           +\displaystyle{1\over18m^6} \biggl[
                  {\cal O}^{\chi_0}_8 ({}^3S^2_1)
                 +5{\cal O}^{\chi_0}_8 ({}^3D^2_1)
					   \biggr].
\end{array}\label{me.chi0me}\end{equation}
The difference between these and the unpolarised cross sections of \cite{upol}
are only in the sign of the coefficient of the $J=2$ components in the
combinations ${\widetilde\Theta}^{\chi_0}_S(9)$ and ${\widetilde
\Theta}^{\chi_0}_D(9)$.

For the ${}^3P_1$ state we find
\begin{equation}
   \Delta \hat\sigma^{\chi_1}_{gg}(\hat s) \;=\;
		\varphi \left[{1\over18}
        \widetilde\Theta^{\chi_1}_S(9) +{5\over48} 
\widetilde\Theta^{\chi_1}_D(9) 
        +{3\over16} \widetilde\Theta^{\chi_1}_F(9)
		     \right],
\label{me.chi1}\end{equation}
where the combinations of non-perturbative matrix elements are,
\begin{equation}\begin{array}{rl}
   \widetilde\Theta^{\chi_1}_S(9)\;=\;&
       {\displaystyle{1\over2m^4}}\left[
               3{\cal O}^{\chi_1}_1({}^3P_0^1)
              -{\displaystyle{4\over5}}{\cal O}^{\chi_1}_1({}^3P_2^1)
                                   \right], \\

   \widetilde\Theta^{\chi_1}_D(9)\;=\;&
        {\displaystyle{1\over2m^2}}{\cal O}^{\chi_1}_8({}^1S_0^0)
       +{\displaystyle{1\over2m^4}}\left[
               3{\cal O}^{\chi_1}_8({}^3P_0^1)
              -{\displaystyle{4\over5}}{\cal O}^{\chi_1}_8({}^3P_2^1)
                                   \right], \\

   \widetilde\Theta^{\chi_1}_F(9)\;=\;& \displaystyle{1\over 6 m^4}  
        {\cal O}^{\chi_1}_8 ({}^1P^1_1) 
       +\displaystyle{1\over3m^6} \biggl[
        \displaystyle{1\over6}{\cal O}^{\chi_1}_8 ({}^3S^2_1)
      \\&\qquad
        +\displaystyle{5\over6}{\cal O}^{\chi_1}_8 ({}^3D^2_1) 
        +\displaystyle{1\over5}{\cal O}^{\chi_1}_8 ({}^3D^2_2) 
					   \biggr].
\end{array}\label{me.chi1me}\end{equation}

The ${}^3P_2$ cross section is
\begin{equation}\begin{array}{rl}
   \Delta \hat\sigma^{\chi_2}_{gg}(\hat s) \;=&\;\varphi
       \biggr[\displaystyle{1\over18}\widetilde\Theta^{\chi_2}_S(5)
             +\displaystyle{1\over18}\widetilde\Theta^{\chi_2}_S(7)
         \\&\qquad\qquad\qquad
             +\left\{\displaystyle{1\over18}\widetilde\Theta^{\chi_2}_S(9)
                    +\displaystyle{5\over48}\widetilde\Theta^{\chi_2}_D(9) 
                    
+\displaystyle{3\over16}\widetilde\Theta^{\chi_2}_F(9)\right\}
		     \biggr]
\end{array}\label{me.chi2}\end{equation}
where the combinations of non-perturbative matrix elements are
\begin{equation}\begin{array}{rl}
   \widetilde\Theta^{\chi_2}_S(5)\;=\;&
        -{\displaystyle{2\over5m^4}}{\cal O}^{\chi_2}_1({}^3P_2^1),\\

   \widetilde\Theta^{\chi_2}_S(7)\;=\;&
      -{\displaystyle{2\over\sqrt{15}m^6}}
                {\cal P}^{\chi_2}_1({}^3P_2^1,{}^3P_2^3),\\

   \widetilde\Theta^{\chi_2}_S(9)\;=\;&
        {\displaystyle{3\over2m^4}}{\cal O}^{\chi_2}_1({}^3P_0^1)
       -{\displaystyle{1\over75m^8}}\biggl[
           {\displaystyle{238\over9}}{\cal O}^{\chi_2}_1({}^3P_2^3)
          +{\displaystyle{141\sqrt3\over2\sqrt7}}
                  {\cal P}^{\chi_2}_1({}^3P_2^1,{}^3P_2^5)
                                  \biggr],\\

   \widetilde\Theta^{\chi_2}_D(9)\;=\;&
        {\displaystyle{1\over2m^2}}{\cal O}^{\chi_2}_8({}^1S_0^0)
       +{\displaystyle{1\over2m^4}}\left[
               3{\cal O}^{\chi_2}_8({}^3P_0^1)
              -{\displaystyle{4\over5}}{\cal O}^{\chi_2}_8({}^3P_2^1)
                                   \right], \\

   \widetilde\Theta^{\chi_2}_F(9)\;=\;& \displaystyle{1\over 6 m^4} 
        {\cal O}^{\chi_2}_8 ({}^1P^1_1) 
        + \displaystyle{1\over3m^6} \biggl[
        \displaystyle{1\over6}{\cal O}^{\chi_2}_8 ({}^3S^2_1) 
        +\displaystyle{5\over6}{\cal O}^{\chi_2}_8 ({}^3D^2_1)
       \\&\qquad
        +\displaystyle{1\over5}{\cal O}^{\chi_2}_8 ({}^3D^2_2) 
        -\displaystyle{2\over7}{\cal O}^{\chi_2}_8 ({}^3D^2_3) 
					   \biggr].
\end{array}\label{me.chi2me}\end{equation}
In the combination ${\widetilde \Theta}^{\chi_2}_S (9)$, the coefficient 
of ${\cal O}^{\chi_2}_1 ({}^3 P^3_2)$ is different from the unpolarised
in magnitude as well as sign, because it arises from several sources. Only
the part from the combination $\Delta{\cal S}\cdot{\cal S}^*$ flips sign.
The other $J=2$ terms change sign as expected. The $J=3$ term in the $F$
colour amplitude is also derived from the $J=2$ part of the $\Delta{\cal S}
\cdot{\cal S}^*$ term, and hence flips sign.

The production cross section for ${}^1S_0$ states is---
\begin{equation}\begin{array}{rl}
   \Delta \hat\sigma^{\eta}_{gg}(\hat s)\;=&\;\varphi
       \biggl[\displaystyle{1\over18}\widetilde\Theta^{\eta}_S(3)
             +\displaystyle{1\over18}\widetilde\Theta^{\eta}_S(5)
       \\&\qquad\qquad\qquad
            +\left\{\displaystyle{1\over18}\widetilde\Theta^{\eta}_S(7)
                   +\displaystyle{5\over48}\widetilde\Theta^{\eta}_D(7) 
                   
+\displaystyle{3\over16}\widetilde\Theta^{\eta}_F(7)\right\}
       \biggr],
\end{array}\label{me.etac}\end{equation}
where the combinations of non-perturbative matrix elements are
\begin{equation}\begin{array}{rl}
   \widetilde\Theta^{\eta}_S(3)\;=\;&
        {\displaystyle{1\over2m^2}}{\cal O}^{\eta}_1({}^1S_0^0),\\

   \widetilde\Theta^{\eta}_S(5)\;=\;&
      {\displaystyle{1\over\sqrt3m^4}}
                {\cal P}^{\eta}_1({}^1S_0^0,{}^1S_0^2),\\

   \widetilde\Theta^{\eta}_S(7)\;=\;&
        {\displaystyle{1\over3m^6}}\left[
            2{\cal O}^{\eta}_1({}^1S_0^2)
           +\displaystyle{4\over\sqrt5}{\cal P}^{\eta}_1({}^1S_0^0,{}^1S_0^4)
                                   \right],\\

   \widetilde\Theta^{\eta}_D(7)\;=\;&
        {\displaystyle{1\over2m^2}}{\cal O}^{\eta}_8({}^1S_0^0),\\

   \widetilde\Theta^{\eta}_F(7)\;=\;& \displaystyle{1\over 6 m^4} 
                          {\cal O}^{\eta_c}_8 ({}^1P^1_1).\\
\end{array}\label{me.etacme}\end{equation}
There is no $J=2$ contribution to $\eta_c$ to ${\cal O} (v^7)$ and hence
the parton level asymmetry is identically equal to 1.

The cross section for the ${}^1P_1$ quarkonium state is---
\begin{equation}\begin{array}{rl}
   \Delta \hat\sigma^{h}_{gg}(\hat s) \;=&\;\varphi
       \biggl[\displaystyle{5\over48}\widetilde\Theta^{h}_D(5)
             +\displaystyle{5\over48}\widetilde\Theta^{h}_D(7) 
         \\&\qquad\qquad\qquad
            +\left\{\displaystyle{1\over18}\widetilde\Theta^{h}_S(9) 
                   +\displaystyle{5\over48}\widetilde\Theta^{h}_D(9)
                   +\displaystyle{3\over16}\widetilde\Theta^{h}_F(9)\right\}
 		     \biggr],
\end{array}\label{me.hc}\end{equation}
where the combinations of non-perturbative matrix elements is given by
\begin{equation}\begin{array}{rl}
   \widetilde\Theta^{h}_D(5)\;=\;&
        {\displaystyle{1\over2m^2}}{\cal O}^{h}_8({}^1S_0^0),\\

   \widetilde\Theta^{h}_D(7)\;=\;&
      {\displaystyle{1\over\sqrt3m^4}}
                {\cal P}^{h}_8({}^1S_0^0,{}^1S_0^2),\\

   \widetilde\Theta^{h}_S(9)\;=\;&
        {\displaystyle{1\over2m^2}}{\cal O}^{h}_1({}^1S_0^0),\\

   \widetilde\Theta^{h}_D(9)\;=\;&
        {\displaystyle{1\over3m^6}}\left[
            2{\cal O}^{h}_8({}^1S_0^2)
           +\displaystyle{4\over\sqrt5}{\cal P}^{h}_8({}^1S_0^0,{}^1S_0^4)
                                   \right]
      \\&\quad
       +{\displaystyle{1\over2m^4}}\left[
               3{\cal O}^{h}_8({}^3P_0^1)
              -{\displaystyle{4\over5}}{\cal O}^{h}_8({}^3P_2^1)
                                   \right]
       +{\displaystyle{1\over15m^6}}{\cal O}^{h}_8({}^1D_2^2),\\

   \widetilde\Theta^{h}_F(9)\;=\;& 
        \displaystyle{1\over 6m^4} {\cal O}^{h}_8 ({}^1P^1_1).
\end{array}\label{me.hcme}\end{equation}
The coefficient of the matrix element ${\cal O}^{h}_8({}^3P_2^1)$
in $\widetilde\Theta^{h}_D(9)$ flips sign as expected. However, the
other $J=2$ term, ${\cal O}^{h}_8({}^1D_2^2)$, is unaffected as it
comes from $\Delta {\cal A}\cdot{\cal A}^*$.

\subsection{$\gamma g\to\bar QQ$ and $\gamma\gamma\to\bar QQ$}

The matrix elements for the two processes $\gamma p\to\bar QQ$ and
$\gamma\gamma\to\bar QQ$ are closely related to the $gg$ amplitudes.
It is easy to check that
\begin{equation}
   {\cal M}_{\gamma g}\;=\; ge D,\qquad{\rm and}\qquad
   {\cal M}_{\gamma\gamma}\;=\; e^2 S,
\label{me.gamma}\end{equation}
where $D$ and $S$ are the colour amplitudes given in eq.\ (\ref{me.ampl}),
and $e$ is the charge of the heavy quark.

The $\gamma g$ cross sections for the production of any quarkonium state
can be obtained from those for the $gg$ process, eqs.\ 
(\ref{me.jpsi})--(\ref{me.hcme}), by the following prescription--- replace
$\alpha_{\scriptscriptstyle S}^2$ in $\varphi$ (eq.\ \ref{me.varphi}) by
$\alpha\alpha_{\scriptscriptstyle S}$, delete the ${\widetilde\Theta}_S$
and ${\widetilde\Theta}_F$ terms, and replace the colour factor $5/48$ for
the terms in ${\widetilde\Theta}_D$ by 2.

These computations may be applied to double polarised asymmetries in
almost elastic photo-production of quarkonium states. Then the hadron level
cross sections are obtained from these parton level quantities by multiplying
the latter by the unpolarised or polarised gluon densities, $g(x)$ and $\Delta
g(x)$ respectively. Here $x=2 m^2/m_{\scriptscriptstyle N}E_\gamma$,
$m_{\scriptscriptstyle N}$ is the proton mass and $E_\gamma$ is the photon
energy in the rest frame of the proton. In phenomenological applications
the diffractive parts of these cross sections have to be separated. Techniques
have been developed at the HERA experiments to do this \cite{hera}.

The $\gamma\gamma$ cross sections are obtained with the
prescription--- replace $\alpha_{\scriptscriptstyle S}^2$ in $\varphi$
(eq.\ \ref{me.varphi}) by $\alpha^2$, delete the ${\widetilde\Theta}_D$
and ${\widetilde\Theta}_F$ terms in eqs.\ (\ref{me.jpsi})--(\ref{me.hcme}),
and replace the colour factor $1/18$ for the terms in ${\widetilde\Theta}_S$
by 16. No parton densities enter for point-like photon cross sections.
If, on the other hand, the cross sections are dominated by resolved photon
processes, then the calculation is the same as for hadro-production.

\section{Phenomenology\label{disc}}

In spite of the large number of non-perturbative matrix elements in the final
results, it is possible to make several quantitative and qualitative comments
about the polarisation asymmetries. The most reliable predictions are
obtained when the parton level asymmetries are $\pm1$ (the ${}^1S_0$ and
${}^1P_1$ quarkonium states) and the matrix elements drop out of the
asymmetries. In these cases the polarisation asymmetries may be used to
measure the polarised gluon densities. In several other cases large
simplifications occur. Some of the asymmetries can be used to test NRQCD
or the scaling laws for matrix elements.

Many of the matrix elements involved in the phenomenology are unknown.
We develop a scaling argument which allows us to make rough estimates.
A dimensional argument, neglecting possible logarithms of $m$ and $v$,
can be used to write
\begin{equation}
   \langle{\cal K}_i\Pi(H){\cal K}^\dagger_j\rangle\;=\;
       R_H Y_{ij} \Lambda^{D_{ij}} v^d,
\label{disc.dimen}\end{equation}
where $D_{ij}$ is the mass dimension of the operator, $d$ is the
velocity scaling exponent in NRQCD (eq.\ \ref{intro.rule}), $\Lambda$ is
the cutoff scale below which NRQCD is defined ($\Lambda\sim m$), and
$Y_{ij}$ and $R_H$ are dimensionless numbers. We have separated these two
factors in order to specify that $Y_{ij}$ contains all trivial numerical
factors as well as Clebsch-Gordan coefficients, for example those which
come from the approximate heavy-quark spin symmetry. $R_H$ contains the
irreducible minimum non-perturbative information. In the context of
potential models, $R_H$ would be related to some integral of the
radial wave function of the state $H$. The scaling rules of NRQCD are a
good guide to the importance of a given term if $R_H$ depends mainly on
$H$ and is approximately independent of the operators ${\cal K}_{i,j}$.

In Table \ref{disc.matel} we collect the values of some of the
non-perturbative matrix elements quoted in the literature and extract from
them the values of $R_H$.
In \cite{bkr} the colour octet matrix elements were fitted to collider
data. We have quoted the values obtained using the GRV LO parton densities.
The good agreement between the values of $R_H$ extracted from the two sets
of octet matrix elements, for both $H=J/\psi$ and $\psi'$, provide some basis
for the working hypothesis that $R_H$ is almost independent of the operator.
In this section we shall accept it, keeping in mind that further checks are
necessary.

\begin{table}\begin{center}\begin{tabular}{|c|c|c|c|}
\hline
Matrix Elements & Values & $R_H$ \\
\hline
${\cal O}^{J/\psi}_8\left({}^3S_1\right)$ &
                  $(1.12\pm0.14)\times10^{-2} {\rm\ GeV}^3$ & $0.22\pm0.03$ \\
${\cal O}^{J/\psi}_8\left({}^1S_0\right)+
            {7\over2m^2}{\cal O}^{J/\psi}_8\left({}^3P_0\right)$ &
                  $(3.90\pm1.14)\times10^{-2} {\rm\ GeV}^3$ & $0.17\pm0.05$\\
\hline
${\cal O}^{\psi'}_8\left({}^3S_1\right)$ &
                  $(0.46\pm0.08)\times10^{-2} {\rm\ GeV}^3$ & $0.09\pm0.02$\\
${\cal O}^{\psi'}_8\left({}^1S_0\right)+
            {7\over2m^2}{\cal O}^{\psi'}_8\left({}^3P_0\right)$ &
                  $(1.60\pm0.51)\times10^{-2} {\rm\ GeV}^3$ & $0.07\pm0.02$\\
\hline
\end{tabular}\end{center}
\caption[dummy]{The values of various non-perturbative matrix elements
  in the non-relativistic normalisation \cite{bkr}. These can be converted
  to the relativistic normalisation by multiplying by $4m$. The dimensionless
  number $R_H$ (eq.\ \ref{disc.dimen}) has been extracted using $\Lambda
  =m=1.5$ GeV, $v^2=0.3$, and $Y=1$, except in the 2nd and 4th lines of
  the table where we used $Y=4.5$.}
\label{disc.matel}\end{table}

Assuming the constancy of $R_H$ and using heavy-quark symmetry, we find
that the non-perturbative matrix elements contribute approximately $12
R_\chi m^2 v^9$ to the $\chi_1$ cross section and about $(1+v^2+12 v^4)
R_\chi m^2 v^5$ to the $\chi_2$ cross sections. Then we expect
\begin{equation}
   {\sigma(\chi_1)\over\sigma(\chi_2)}\;\approx\;
        {12v^4\over1+v^2+12v^4}\;=\;0.45,
\label{disc.rat}\end{equation}
independent of $\sqrt S$. This estimate\footnote{The number of terms
allowed at higher orders in $v$ can grow at most by 12 for each order.
If the coefficient functions continue to be of order unity, then the all
orders result for the $\chi_1/\chi_2$ ratio can be bounded by $0.67$ for
$\chi_c$ and by $0.12$ for $\chi_b$. The ratio in photo-production is close
to unity in both these cases. In \cite{br} an estimate of ${\cal O}
(\alpha^3_{\scriptscriptstyle S})$ effects was used to show that the
$\chi_1/\chi_2$ ratio could be about $0.3$.} is in reasonable agreement with
the measured values in proton-nucleon collisions--- $0.34\pm0.16$ at $\sqrt
S=38.8$ GeV \cite{e771} and $0.24\pm0.28$ at $\sqrt S=19.4$ \cite{e673}. The
measurements are also compatible with a lack of $\sqrt S$ dependence. In
NRQCD this ratio cannot depend on the beam hadron. It turns out that the
estimate in eq.\ (\ref{disc.rat}) is not very far from the recently measured
value in pion-nucleon collisions--- $0.57\pm0.19$ at $\sqrt S=31.1$ GeV
\cite{e706}. However, the experimental situation certainly needs
clarification.

\begin{figure}
\vskip6truecm
\includegraphics{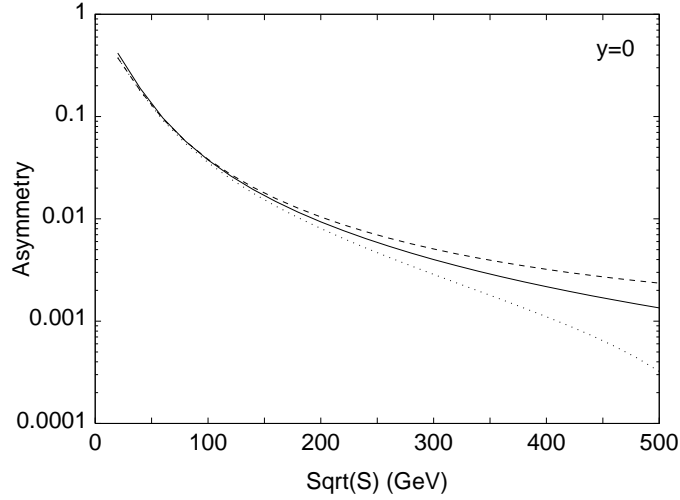}
\caption[dummy]{The asymmetries (a) $\Delta{\cal L}_{gg}/{\cal L}_{gg}$
  (full line) (b) $(\Delta{\cal L}_{gg}+\Delta{\cal L}_{\bar qq})/
  ({\cal L}_{gg} +{\cal L}_{\bar qq})$ (dashed line) and (c) 
  $(\Delta{\cal L}_{gg}-\Delta{\cal L}_{\bar qq})/({\cal L}_{gg}
  +{\cal L}_{\bar qq})$ (dotted line) for zero rapidity as a function of
  the CM energy $\sqrt S$ using the GRSV LO parton density set \cite{grv}. 
  The largest asymmetries are obtained at the lowest values of $\sqrt S$,
  where all the asymmetries are dominated by the gluon asymmetry.}
\label{fg.asym}\end{figure}

This paper is concerned with low transverse momentum production. Our results
are best suited to help design fixed target polarised experiments, for example
by extracting the RHIC or HERA polarised beams for use on polarised targets.
The reason is that the polarisation asymmetries are likely to be largest
at small $\sqrt S$ (Figure \ref{fg.asym}), yielding the best signals. Also,
at smaller values of $\sqrt S$ the $\bar qq$ channel seems to be unimportant
for the phenomenology, and the polarised gluon asymmetries can be studied
directly. Since many aspects of low-energy quarkonium production can be
tested by such experiments, it is worth considering them seriously.

For ${}^1S_0$ quarkonium production, the parton level asymmetry in the $gg$
channel is identically 1. For the yet to be discovered ${}^1P_1$ resonances,
this is true with corrections of order $v^4$ (from the
${\widetilde\Theta}^{h}_D(9)$ term in eq.\ \ref{me.hcme}). Only the colour
amplitude $S$ enters in $\gamma\gamma$ reactions; hence the polarisation
asymmetry is exactly 1 for both these states. In $\gamma p$ collisions 
\begin{equation}
   A_{\gamma p}\;=\; {\Delta g(x)\over g(x)}
               \qquad\qquad({}^1S_0,\;{}^1P_1{\rm\ quarkonia}),
\label{disc.gpeta}\end{equation}
For ${}^1S_0$ states the above equation is exact, whereas there are
order $v^4$ corrections for ${}^1P_1$ states.

For $pp$ collisions, neglecting the $\bar qq$ channel for reasons explained
earlier, we find that
\begin{equation}
   A_{pp}\;=\; \left[1+{\cal O}(v^4)\right]
         {\Delta{\cal L}_{gg}\over{\cal L}_{gg}}
               \qquad\qquad({}^1S_0,\;{}^1P_1{\rm\ quarkonia}),
\label{disc.ppeta}\end{equation}
where the order $v^4$ corrections come from the $\bar qq$ channel for the
${}^1S_0$ resonance, and in addition from the order $v^9$ matrix elements
for the ${}^1P_1$ resonance.

The case of ${}^3P_J$ states is very interesting phenomenologically. For
$\gamma p\to\chi_{\scriptscriptstyle J}$ reactions, we need to retain
only the contribution from the colour amplitude $D$. The cross sections
then begin at order $v^9$, and we find that
\begin{equation}
   A_{\gamma p}\;=\; {\Delta g(x)\over g(x)}\,
      {{\widetilde\Theta}_D^{\chi_J}(9)\over\Theta_D^{\chi_J}(9)}
               \qquad\qquad({}^3P_J{\rm\ quarkonia}).
\label{disc.gpchi}\end{equation}
Since both the non-perturbative matrix elements involved are unknown,
it might seem that there is no predictive power to the model. However,
using heavy quark symmetry, and the scaling form of eq.\ (\ref{disc.dimen})
it appears that ${\widetilde\Theta}_D^{\chi_J}(9)=0$. As a result, we obtain
the prediction
\begin{equation}
   A_{\gamma p}^{\chi_0}\;=\; 
   A_{\gamma p}^{\chi_1}\;=\; 
   A_{\gamma p}^{\chi_2}\;\approx\;
      [0+{\cal O}(v^2)] {\Delta g(x)\over g(x)},
\label{disc.chiphen}\end{equation}
where the order $v^2$ terms come from the un-calculated order $v^{11}$ terms
in the asymmetry and from corrections to heavy quark spin symmetry.

We can make similar arguments for the polarisation asymmetries in
$\gamma\gamma$ collisions. We find that
\begin{equation}
   A_{\gamma\gamma}^{\chi_0}\;=\; 1 +{\cal O}(v^4) \;=\;
      - A_{\gamma\gamma}^{\chi_2}.
\label{disc.ggchi}\end{equation}
The case of $\chi_1$ is more complicated, since a ratio of two unknown
matrix elements is involved. However, using heavy-quark spin symmetry
this reduces to
\begin{equation}
   A_{\gamma\gamma}^{\chi_1}\;=\; -{1\over7} +{\cal O}(v^2).
\label{disc.ggchip}\end{equation}
Unfortunately, no polarised $\gamma\gamma$ experiments have been planned
to test this simple prediction of NRQCD and heavy quark spin symmetry.

A straightforward application of the NRQCD scaling laws would lead us
to the conclusion that the asymmetries for $pp\to\chi_{0,2}$
are given by
\begin{equation}
   A_{pp}^{\chi_0}\;\approx\;
   -A_{pp}^{\chi_2}\;\approx\;
       {\Delta{\cal L}_{gg}\over{\cal L}_{gg}}+{\cal O}(v^4),
\label{disc.ppchi02}\end{equation}
where we have neglected the contribution of the $\bar qq$ channel. The
asymmetry for $\chi_1$ production is
\begin{equation}
   A_{pp}^{\chi_1}\;=\;
     {{\widetilde\Theta}_S^{\chi_1}(9) + {\widetilde\Theta}_D^{\chi_1}(9)
                                   + {\widetilde\Theta}_F^{\chi_1}(9)
     \over
     \Theta_S^{\chi_1}(9) + \Theta_D^{\chi_1}(9) + \Theta_F^{\chi_1}(9)}
       \left[{\Delta{\cal L}_{gg}\over{\cal L}_{gg}}\right].
\label{disc.ppchip}\end{equation}
The ratio of matrix elements can be estimated using heavy quark spin
symmetry and the scaling relations in eq.\ (\ref{disc.dimen}).
${\widetilde\Theta}_D^{\chi_J}(9)$ vanishes in this approximation and
the terms ${\widetilde\Theta}_{S,F}^{\chi_J}(9)$ come with opposite
signs. The numerator is positive but small and we expect---
\begin{equation}
   A_{pp}^{\chi_1}\;\approx\;
       0.2\,{\Delta{\cal L}_{gg}\over{\cal L}_{gg}}.
\label{disc.ppchi1}\end{equation}
The $\bar qq$ channel remains negligible even at $\sqrt S=500$ GeV.

The $\psi'$ asymmetry is straightforward. If the $\bar qq$ channel is
neglected then
\begin{equation}
   A^{\psi'}_{pp}\;=\;{\Delta{\cal L}_{gg}\over{\cal L}_{gg}}\,
     {\widetilde\Theta^{\psi'}_D(7)\over\Theta^{\psi'}_D(7)}
        +{\cal O}(v^2),\quad
   A^{\psi'}_{\gamma p}\;=\;{\Delta g(x)\over g(x)}\,
     {\widetilde\Theta^{\psi'}_D(7)\over\Theta^{\psi'}_D(7)}
        +{\cal O}(v^2),
\label{disc.allpsi}\end{equation}
and the form of the non-perturbative matrix elements is the same as
that for $J/\psi$.
The numerator can be shown to vanish using heavy quark spin symmetry
and eq.\ (\ref{disc.dimen}). A numerical estimate of the subleading
order $v^9$ terms shows that the ratio of non-perturbative matrix
elements is small (of the order of 0.05). At low $\sqrt S$ the
asymmetry is dominated by the asymmetry in the gluon densities.
However, at $\sqrt S$ of a few hundred GeV, the gluon asymmetry
reduces to about 0.001, and the neglected $\bar qq$ channel becomes
important. These terms appear with the opposite sign and reduce the
asymmetry even further. As a consequence, we expect the $\psi'$
asymmetry to be small for all $\sqrt S$---
\begin{equation}
   A^{\psi'}_{pp}\;\approx\;0,\qquad
   A^{\psi'}_{\gamma p}\;\approx\;0.
\label{disc.psip}\end{equation}
Any attempt to see structure in this asymmetry would probably require
unreasonably large statistics.

The $J/\psi$ asymmetry seems to be enormously complicated because of
the feed-down from radiative decays of the $\chi$ states. However, a
major simplification occurs because of the near vanishing asymmetry
in direct $J/\psi$ production (the argument is the same as for $\psi'$).
Thus the asymmetry comes entirely from the 20--40\% of the cross section
due to $\chi$ decays. Taking into account the ratios of the production
cross sections of $\chi$ and the branching fractions for their decays
into $J/\psi$, we find that the $\chi_1$ and $\chi_2$ states contribute
equally to $J/\psi$. Hence the $J/\psi$ polarisation asymmetry is
expected to be approximately
\begin{equation}
   A_{pp}^{J/\psi}\;\approx\; -(0.15\pm0.05)
       {\Delta{\cal L}_{gg}\over{\cal L}_{gg}}.
\label{disc.psi}\end{equation}
We summarize the predictions made on the basis of the NRQCD scaling in
eq.\ (\ref{disc.dimen}) and the assumption of $R_H$ depending only on
the hadron $H$---
\begin{equation}\begin{array}{rl}
   A_{pp}^{\psi'}\;\ll\;
   -A_{pp}^{J/\psi}\;\approx\;
   A_{pp}^{\chi_1}\;<\;
   A_{pp}^{\chi_0}\;=\;
   -A_{pp}^{\chi_2}\;=\;
       {\Delta{\cal L}_{gg}\over{\cal L}_{gg}},\\
   A_{\gamma p}^{\psi'}\;\approx\;
   A_{\gamma p}^{J/\psi}\;\approx\;
   A_{\gamma p}^{\chi_1}\;\approx\;
   A_{\gamma p}^{\chi_0}\;\approx\;
   A_{\gamma p}^{\chi_2}\;\approx\; 0.
\end{array}\label{disc.nrqcd}\end{equation}
The asymmetry for $J/\psi$ in almost-elastic photo-production vanishes
because the direct as well as feed-down contributions vanish.

Alternative points of view can be taken about the scaling of the
non-perturbative matrix elements. These proceed on the assumption that
the mass scale, $\Lambda$, appearing in the soft matrix elements are
smaller than $m$. If the scale $\Lambda$ in eq.\ (\ref{disc.dimen}) is
taken to be the QCD scale $\Lambda_{QCD}$ \cite{upol}, then
the relation $v\sim\alpha_{\scriptscriptstyle S}(mv)$ \cite{bbl} implies
that $z\equiv\Lambda_{QCD}^2/m^2<v^4$ for all $m$. The relative importance
of terms then changes dramatically, since the lowest dimensional operators
dominate. Since the dimension 4 terms ${\cal O}^{\chi_J}_8({}^1S_0^0)$ appear
in the $\chi$ cross sections at order $v^9>zv^5$, they dominate the cross
section. Heavy quark spin symmetry then gives the $\chi_1/\chi_2$ ratio
to be $5/3$, in agreement with the pion-nucleon observations.
The same $\chi_1/\chi_2$ ratio is then predicted for charm and bottom
mesons in hadro- and photo-production, and the result remains stable
even when higher order terms are introduced. The rearrangement of
the series also leads to
$\Theta^{J/\psi}_D(7)\approx{\widetilde\Theta}^{J/\psi}_D(7)\ne0$ and
$\Theta^{\chi_J}_D(9)\approx{\widetilde\Theta}^{\chi_J}_D(9)\ne0$.
The consequences for the asymmetries are---
\begin{equation}\begin{array}{rl}
   A_{pp}^{J/\psi}\;\approx\;
   A_{pp}^{\psi'}\;\approx\;
   A_{pp}^{\chi_0}\;\approx\;
   A_{pp}^{\chi_1}\;\approx\;
   A_{pp}^{\chi_2}\;=\;&
       {\displaystyle{\Delta{\cal L}_{gg}\over{\cal L}_{gg}}},\\
   A_{\gamma p}^{J/\psi}\;\approx\;
   A_{\gamma p}^{\psi'}\;\approx\;
   A_{\gamma p}^{\chi_0}\;\approx\;
   A_{\gamma p}^{\chi_1}\;\approx\;
   A_{\gamma p}^{\chi_2}\;=\;&
       {\displaystyle{\Delta g(x)\over g(x)}},
\end{array}\label{disc.alt}\end{equation}
and totally different from the NRQCD results summarized in eqs.\ 
(\ref{disc.nrqcd}).
These differences make low-energy double polarised experiments a good
testing ground for NRQCD, and the entire phenomenology of low-energy
quarkonium production. Other choices of the mass scale $\Lambda$ have
also been suggested \cite{schuler}. These give somewhat different
results. The asymmetries for $\eta_{c,b}$ and $h_{c,b}$ are independent
of such assumptions, and hence form good measures of polarised gluon
densities.

In conclusion, low energy double polarised asymmetries are a good test-bed
for understanding the origin of all observed systematics in fixed target
hadro- and photo-production of charmonium. The high order computations
presented here provide us with a set of processes which can be used to test
aspects of NRQCD factorisation and scaling. In addition, ${}^1S_0$ and
${}^1P_1$ charmonium and bottomonium states are well suited for measurements
of polarised gluon densities.
Assuming standard NRQCD power counting laws, higher order terms are likely
to be phenomenologically important for double polarisation asymmetries of
${}^3P_1$ and excited ${}^3S_0$ states in both hadro- and photo-production.

\end{document}